# Challenges: Bridge between Cloud and IoT

Mohammad Riyaz Belgaum[1,3], Safeeullah Soomro[1],
Zainab Alansari[1,2]

[1]Department of Computer Studies
AMA International University, Kingdom of Bahrain.
[2]Department of Computer Science & Information Technology
University of Malaya, Kuala Lumpur, Malaysia.
{ bmdriyaz, s.soomro & zeinab }@ amaiu.edu.bh

Shahrulniza Musa[3]

Malaysian Institute of Information Technology
3Universiti Kuala Lumpur (UniKL MIIT),
Kuala Lumpur, Malaysia

shahrulniza@unikl.edu.my

Muhammad Alam[4]

[4]College of Computer Science and IS (IOBM) Karachi, Pak
Institute of Post Graduate Studies

Universiti Kuala Lumpur (UniKL IPS), KL Malaysia

malam@iobm.edu.pk

Mazliham Mohd Su'ud[5]

5Malaysia France Institute
Universiti Kuala Lumpur, (UniKL MFI),
Bandar Baru Bangi, Malaysia

mazliham@unikl.edu.my

*Abstract* – **In the real time processing, a new emerging technology where the need of connecting smart devices with cloud through Internet has raised. IoT devices processed information is to be stored and accessed anywhere needed with a support of powerful computing performance, efficient storage infrastructure for heterogeneous systems and software which configures and controls these different devices. A lot of challenges to be addressed are listed with this new emerging technology as it needs to be compatible with the upcoming 5G wireless devices too. In this paper, the benefits and challenges of this innovative paradigm along with the areas open to do research are shown.**

## I. INTRODUCTION

Cloud computing is an accelerating technology that allows to use the IT capabilities according to the user's desire or according to the business needs and can be used anywhere – at office, house or any other holiday site for the reason that the cloud is accessible by a network like the Internet. It provides 'as a service' to IT capabilities, such as software applications, storage, network, interface, infrastructure etc.

Internet of Things is a new concept in information technology world and communication sectors [1]. In short, it is a modern technology where gives every creature such as humans, animals or objects the ability to send and receive data via network communication ranging from the Internet or an intranet [2]. Smart devices are gathered in one batch which is called Internet of Things. At a basic level, in fact, Internet of Things assist on the relationship between different objects via the Internet and their communication with each other to achieve the purpose of providing more efficient and more intelligent experience [3]. As with any new technology, IoT seemed a confusing concept at first. Moreover, this concept defined new and unique meanings especially when it comes to safety and security standards [4]. In other words, the idea of designing different devices with an ability to have a wireless communication to be tracked and controlled via the Internet or even through a single Smartphone's application describe the term: Internet of Things [5].

*Benefits of merging Cloud with IoT are as follows:*

*Affordable:* Customers do not have to make any investments in their infrastructure or equipment, so cloud proves very economical in the long run. In addition, the payment of Cloud is, through the 'pay as you go' method depending on the demand of the needed resources, to the service provider. In this way the customer avoids unnecessary or extra payments and only gives the same amount as the resources are used by the business - fewer resources and less resources and more money.

*Secure:* If your laptop or business phone is lost, then the critical business data goes along with it and it does not lessen any terrible tragedy. But if the data is stored in the cloud like OneDrive, then customer does not have to fear it, because it can be accessed by using any device or machine and remotely can wipe all critical data from the lost device.

*Efficient growth:* The optimum capacity of resources in the cloud is used, so the cloud solutions prove to be very beneficial. Cloud not only enhances the efficiency of the business but also increases the efficiency of the employees, because the use of cloud applications can be used by employees not just in the office but anywhere - all of their mails, documents, everything, so with the help of the cloud right In effect, 'office on the go' is realized.

*Elastic and scalable:* Whether it is bandwidth, storage or any other resource, everything can be enhanced and reduced according to the business demand in the cloud. It is much better than the old ways of hosting where the prescribed resources are given and whether you use them or not, you have to pay the full amount.

*Avoiding Downtime or Delay:* Cloud is a network of connected servers. Therefore, if any node fails, then all of its load is taken away by the second cloud node, so the business site is on the cloud server, or cloud services, is never down.

*Disaster recovery:* Big business can afford additional IT resources for disaster management, but this is an additional expense for SMBs because IT resources used for disaster management are left empty, as they will be used only at the time of that disaster. Cloud SMBs provide a low cost and safe disaster recovery mechanism.

*Environmental compatibility:* Cloud conserves energy and provides efficient technical solutions to the business by emitting less carbon percentage than efficient use of resources resulting in Green Computing.

*Incentives to new experiments*: For a developer, or a tester or IT engineer, it gives scope to test your experiments easily with the cloud - harmless live environment and these cloud test infrastructure which will be changed frequently according to testing. It can be very economical compared to the change in the real environment.

It can be clearly stated that the Internet of Things has dominated in many different areas and is to be formally recognized [6]. The IoT potential application areas which played a vital role and gained increasing attentions includes smart cities (smart zones), smart cars, smart houses, smart health, smart industries, public safety, energy and environmental protection, agriculture and tourism [7]. Most of the governments in Europe, America, and Asia considered the Internet of Things as an example and a symbol of innovation and growth [8]. Although great actors still do not recognize some potential of its software applications, many have given up or even invented and created new conditions for the Internet of Things and are adding additional components to it [9]. Furthermore, today, end users and enterprises in the private sphere, have gained considerable expertise in dealing with smart devices and network applications [10].

Overcoming these obstacles, could be the result of better potential utilization of the Internet of Things which provides a stronger interoperability, raise awareness in the real world and use a problem-solving infinite space [11]. According to Gartner 2017[12], in Fig 1, shown below, listed are some of the trends which are emerging and will be obsolete in due course of time with the time span mentioned. Out of the broadly classified categories, IoT in the field of digital platform is at the peak showing its pace.

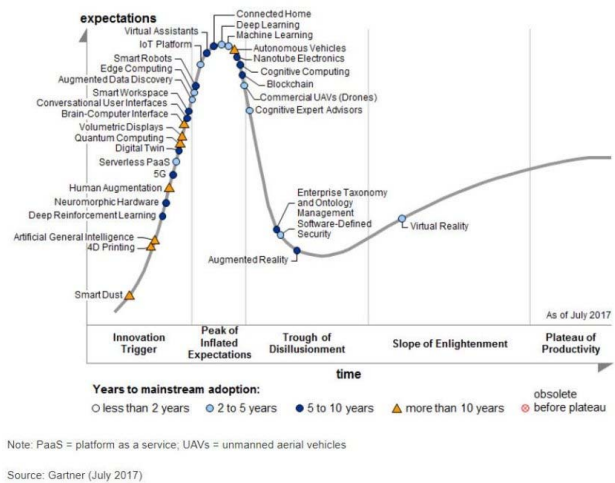

Fig 1 : Hype Cycle for Emerging Technologies, 2017[12].

## II. RELATED LITERATURE

In real time while scheduling the applications for processing received from the smart devices using cloud computing technology, scheduling algorithms are followed. Some of the Load Balancing algorithms have been examined in detail, like one of them being implemented on checkpoint based [13]. The cloud services were ranked based on the checkpoint based load balancing considering the client's prerequisites and keeping up the QoS. But the ranking of services cannot be applied to all the type of devices as each device has a different configuration in real world.

The authors in [14] have examined on how cloud computing applications have broadened their services in mix with portable and quick moving correspondence media know as mobile cloud computing. The structures, qualities, similitudes, and difficulties identified with both of the area have been examined there. Further, in [15], the authors have demonstrated how HTML5 is utilized to execute the applications and services of cloud in a proficient way. All things considered, the holes between conventional cloud computing and versatile cloud computing were appeared.

In [16], the authors discussed on the channel utilization of the devices connected with different IP on a mesh topology in NoC architecture. The proposed calculative utility channel can be applied to different IP's with an internal port address of 4 bit.

All computers connected to the Internet can connect and communicate with each other. Internet development is based on information level and social communication. With IoT all objects that surrounded us are connecting to each other via internet. Communication in IoT is much more than M2M, wireless sensor networks, 2G / 3G / 4G, RFID and etc[17]. In network neutrality field, it is an essential element where no bits of information have priority over the other. Thus, based on the connection principles, anything from / to anyone is located anywhere and is using the most appropriate available physical path between transmitter and receiver at any time.

The authors in [18] clarified the differences between Internet of Things (IoT), Internet of Everything (IoE) and Internet of Nano Things (IoNT), as to make it clear in using the terms based on the characteristics. Current advancements happening in each of the field have been explained with the areas left for the researchers to find solutions in the field.

In the field of Healthcare the authors in [19] explained about how the data is translated into a unified format. Cryptography being not always possible on types of data, and not on all types of channels of communications which can be moved to clouds, a new technique called statistical fingerprinting was introduced. And this technique was used to transfer the sensor data by switching from non HL7 based non secure Health Information System to secured HIS.

IoT in the healthcare has been studied in [20] and FAHP research methodology is used to rank the benefits of using IoT in healthcare. The sub categories like Quality of Life, Environmental protection and Economic prosperity were considered and weights were given from the survey data collected. On each of the criteria used it showed the priorities so that the policy makers can focus on the technologies to improve them to better serve in the area of healthcare.

## III. DISCUSSION

The combination of the Internet of Things is estimated with methods of related technologies and concepts such as Cloud Computing, future Internet, Big data, Robotics and Semantic Technology [21]. Internet of Things potential is still developing and growing about some factors that have restricted the full advantage of the IoT such as:

- Quality of service [22] is a very important factor which is measured based on the bandwidth, the processing speed and the service itself which it provides. There is no clear global scale strategy is lasting or permanent for the use of unique ID or indexed spaces of different object's types.
- There is no acceleration, momentum and further development of the IoT reference architecture such as Architecture Reference Model (ARM) for Internet of Things projects [23].
- There has been limited progress in semantic interoperability for the sensor data exchange in heterogeneous environments [24].
- There are some problems in developing a clear innovation approach, information trust and ownership on Internet of Things, as simultaneously maintaining the security and privacy in an environment is a complex task [25].
- There are some problems in business development that embraces the full potential of the Internet of Things [26].
- Large-scale testing and learning environments, which facilitate testing with complex sensor networks and through feedback and experience lead to innovation, have been achieved in the lower levels [27].
- Only a small amount of rich user interfaces has been developed and need attention in the integration areas [28] as the new generation devices are emerging belonging to 5G.

Practical aspects have not been fulfilled. Such as remarkable roaming charges for sensor applications in a geographical wide range, the unavailability of technical and Lack of reliable network connection are among the raised issues [29]. Moreover as these tow technologies merge, the challenges too have been doubled, so as to address each of them individually and when are combined the new challenges comes need to be addressed.

The protocols to communicate between devices in connection oriented or connectionless environment provide secure transmission like in mobile adhoc networks [30]. But there is no standard protocol used in Cloud of things to communicate as the devices are of different configurations and each device transfer the data in various formats.

In this context, the network convergence concept using IP is essential and relies on the use of multiple services of common IP network and supporting a broad range of applications and services. Fig 2 shows that how the IP is used to communicate and control small devices, sensors, real time IT-oriented networks and specialized networking programs ease the way of having full convergence.

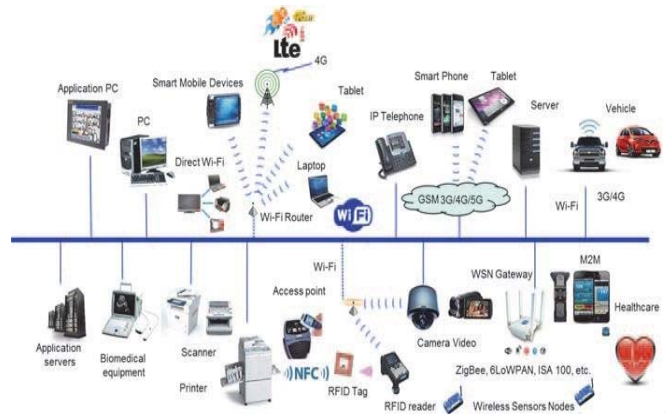

Fig 2 : IP convergence to communicate and control the devices.[31]

Currently, the Internet of Things are composed of heterogeneous loose sets which are in fact, specific purpose networks without the internal connections. For example, modern vehicles have several systems to control the engine performance, safety features, and communication systems and so on. Commercial and residential buildings also have various control systems for heating, ventilation, air conditioning (HVAC), telephone services, security, and lighting. The Internet of Things revealed with high security, analysis, and management capabilities and some of them can be merged. These conditions will allow the IoT to become more powerful and approach a position where it can help more people. Fig 3 shows the presence of the IoT as a network of networks.

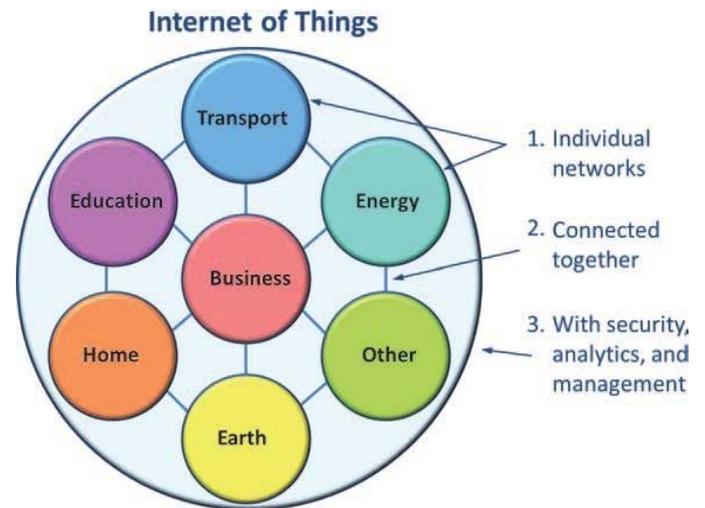

Fig 3 : IoT as a network of networks. (Source: Cisco IBSG, April 2011)

The Internet of Things is not a single technology but a concept in which many new objects are connected and activated. For example the street lights that are networked together, objects with embedded sensors, image recognition capabilities, augmented reality, close field communication with decision in position, management of new resources and services and etc. These have created many business opportunities and added the complexity of information technology. Distribution, transportation, procurement, reverse logistics, service environment are all areas in which information and "objects" are connected to each other and produced new business processes or they have created a much more efficient and profitable inventory.

The IoT provides aggregate IT-based solutions, which refers to the usage of hardware and software to store, retrieve, process data and technology communication including an electronic system for communication between individuals or groups. The rapid convergence of information technology and communication technology are taking place in three pillars of technological innovation. These three layers which are shown in Fig 4 include cloud, data, and communications of pipes, networks, and devices.

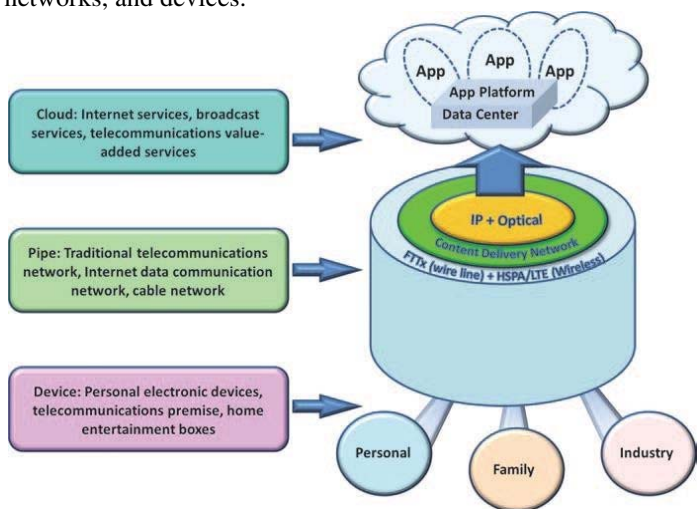

Fig 4 : Three Layers in Technological Innovation.[31]

Collaboration on accessing and exchanging of potential information provides new opportunities for the internet of things applications. Currently, in over 70% of the companies, all objects which are connected to the internet are monitored only through an executable file. It is predicted that by 2020, there will be more than 30 billions of connected objects with over 200 billion intermittent connections. Furthermore, in creating the ability to use a large number of objects that are connected to the IoT, cognitive technologies and underlying intelligence play a significant role. The internet is not just a network of computers, but the internet on a network of objects with any kind and size, vehicles, smart phones, home appliances, toys, cameras, medical tool and industrial systems, which are all connected to the world wide web and communicate and share information at all times.

The Internet of Things has had different meanings as a service providers had a level lower than average. Today, the IoT has become a "universal concept" which requires a standard definition. Taking into account, the significant, extensive history and technologies such as measuring devices, subsystem communications, data collections and initial processing to create object samples and finally, the provision of services, producing an unambiguous definition of "internet of things" is important and non-obvious.

European Research Cluster on the internet of things (IERC) definition of IoT is a global infrastructure for the intelligence community, providing advanced services by connecting objects (physical and virtual), compatible information and evolving and communicating technology.

IV. CONCLUSION

Internet of Things continues to prove its important position in IT and communications and further development in the community. While the basic concepts and foundations are carefully described and reached maturity, greater efforts are needed to unleash its full potential and consolidate and federate systems and actors in the use of its facilities in various fields. The benefits and challenges are going hand in hand with this new innovative technology however to better serve the community all the challenges are to be addressed.

V. Future Studies

Undoubtedly, this research can be the basis for the researchers to address the challenges while these two widely accepted technologies are merged.